\documentclass[preprint]{sigplanconf}

\usepackage{amsmath}
\usepackage{graphicx}
\usepackage{hyperref}
\usepackage{url}

\newcommand{\MATLAB}{\textsc{MATLAB}}
\newcommand{\Mathematica}{\textit{Mathematica}}
\newcommand{\code}[1]{\texttt{#1}}

\begin{document}

\setlength{\pdfpageheight}{\paperheight}
\setlength{\pdfpagewidth}{\paperwidth}

\conferenceinfo{ARRAY '14}{June 15, 2014, Edinburgh, UK}
\copyrightyear{2014} 
\copyrightdata{978-1-nnnn-nnnn-n/yy/mm} 
\doi{nnnnnnn.nnnnnnn}

% Uncomment one of the following two, if you are not going for the 
% traditional copyright transfer agreement.

\exclusivelicense                % ACM gets exclusive license to publish, 
                                  % you retain copyright

%\permissiontopublish             % ACM gets nonexclusive license to publish
                                  % (paid open-access papers, 
                                  % short abstracts)

\titlebanner{Array Operators Using Multiple Dispatch}    % These are ignored unless
\preprintfooter{Array Operators Using Multiple Dispatch} % 'preprint' option specified.

% \title{Array implementations in Julia}

%\title{Array Operators Using Multiple Dispatch}
%\subt
\title{ Array Operators Using Multiple Dispatch }
\subtitle{A design methodology for array implementations in dynamic languages}

\authorinfo{Jeff Bezanson \and Jiahao Chen \and Stefan Karpinski \and Viral Shah  \and Alan Edelman}
           {MIT Computer Science and Artificial Intelligence Laboratory}
  { bezanson@mit.edu, jiahao@mit.edu, stefan@karpinski.org, viral@mayin.org, edelman@mit.edu }

\maketitle

\begin{abstract}

Arrays are such a rich and fundamental data type that they tend to be built into
a language, either in the compiler or in a large low-level library.
Defining this functionality at the user level instead provides greater
flexibility for application domains not envisioned by the language designer.
Only a few languages, such as C++ and Haskell, provide the necessary power to define
$n$-dimensional arrays, but these systems rely on compile-time abstraction,
sacrificing some flexibility.
In contrast, dynamic languages make it straightforward for the user to define any
behavior they might want, but at the possible expense of performance.

As part of the Julia language project, we have developed an approach that yields
a novel trade-off between flexibility and compile-time analysis. The core
abstraction we use is multiple dispatch.
We have come to believe that while multiple dispatch has not been especially popular
in most kinds of programming, technical computing is its killer application.
%This has been extensively studied
%as a general object-oriented programming technique, but we find it
%especially relevant to technical and array-based programming.
By expressing key functions such as array indexing using multi-method
signatures, a surprising range of behaviors can be obtained, in a way that is
both relatively easy to write and amenable to compiler analysis.
The compact factoring of concerns provided by these methods makes it easier
for user-defined types to behave consistently with types in the standard
library.

%pulls out abstractions that might not have been named before
%creates better integration of user-defined types
%flexible enough to change the behavior if you want
%creates more coherence

%Interest in recovering performance in these systems has spurred development of
%analysis frameworks. 

\end{abstract}

%\category{CR-number}{subcategory}{third-level}


\section{Array libraries}

\begin{quotation}
``Unfortunately, it is very difficult for a designer to select in advance all
the abstractions which the users of his language might need. If a language is
to be used at all, it is likely to be used to solve problems which its
designer did not envision, and for which abstractions embedded in the language
are not sufficient.'' - Ref. \cite{Liskov:1974pb}
\end{quotation}

$n$-arrays (arrays of rank $n$, or simply arrays) are an essential data
structure for technical computing, but are challenging to implement
efficiently \cite{Sattley:1960as,Sattley:1961as,Randell:1964a6}. There is a
long history of special-purpose compiler optimizations to make operations over
arrays efficient, such as loop fusion for array traversals and common
subexpression elimination for indexing operations \cite{Randell:1964a6,
Busam:1969oe}. Many language implementations therefore choose to build array
semantics into compilers.

Only a few of the languages that support $n$-arrays, however, have sufficient
power to express the semantics of $n$-arrays for general rank $n$ without
resorting to hard-coding array behavior into a compiler.
Single Assignment C \cite{Grelck:2006sa} is a notable language with built-in
$n$-array support. Other languages
have well-established array libraries, like the C++ libraries 
\code{Blitz++} \cite{Veldhuizen:1998ab} and \code{Boost.MultiArray}
\cite{Garcia:2005ma} and Haskell's \code{Repa} (Regular Parallel Arrays)
\cite{Keller:2010rs,Lippmeier:2011ep, Lippmeier:2012gp}. These libraries
leverage the static semantics of their host languages to
define $n$-arrays inductively as the outer product of a 1-array with an
$(n\!\!-\!\!1)$-array \cite{Bavestrelli:2000ct}.
Array libraries typically handle dimensions recursively, one at a time;
knowing array ranks at compile-time allows the compiler to infer the amount
of storage needed for the shape information, and unroll index computations fully.

%The dispatch semantics also allows for efficient
%implementations of certain operations; for example, \code{Repa} allows for in-
%memory transposition by dispatching on classes with different memory striding
%rules \cite{Keller:2010rs}. The use of compile-time features allows these
%libraries to eliminate runtime overhead for better performance.

\subsection{Static tradeoffs}

Array libraries built using compile-time abstraction have good performance,
but also some limitations.
First, language features like C++ templates are not available at run-time, so these
libraries do not support $n$-arrays where $n$ is known only at run-time.
Second, code using these features is
notoriously difficult to read and write; it is effectively written in a
separate sublanguage.
Third, the recursive strategy for defining $n$-arrays
naturally favors only certain indexing behaviors. For example,
\code{Repa}'s reductions like \code{sum} are only defined naturally over the
last index \cite{Keller:2010rs}; reducing over a different index requires
permutations.
However, it is worth noting that Haskell's type system encourages
elegant factoring of abstractions. While the syntax may be unfamiliar,
we feel that \code{Repa} ought to hold much interest for the technical computing
community.
%permutation operations and are not guaranteed to produce an optimal memory
%traversal pattern. The overhead incurred by general indexing operations can be
%avoided only by implementing many special cases that ensure maximal
%exploitation of loop unrolling optimizations and other similar static analyses
%\cite{Garcia:2005ma, Lippmeier:2011ep}, which engenders much repetition in the
%codebase \cite{Lippmeier:2012gp}.

Some applications call for semantics that are not amenable
to static analysis. Some may require arrays whose ranks are known only at
run-time, e.g. when reading in arrays from disk or from a data stream where the
rank is specified as part of the input data. In these programs, the data
structures cannot be
guaranteed to fit in a constant amount of memory. Others may
wish to dynamically dispatch on the rank of an array, a need which a
library must anticipate by providing appropriate virtual methods.

%different semantics for arrays of different ranks. Some technical computing
%applications, for example, wish to treat vectors (1-arrays) differently from
%matrices (2-arrays), supporting computations such as the singular value
%decomposition for the latter but not the former. In such contexts, column
%vectors are no longer equivalent to $N\times1$ matrices or 2-arrays with a
%trailing singleton dimension, and programs that conflate them violate the
%Liskov substitution principle \cite{Liskov:1987da}, in ways are are analogous
%to the canonical circle-ellipse problem \cite{Halbert:1987ut}.

\subsection{Dynamic language approaches}

Applications requiring run-time flexibility are better expressed in
dynamic languages such as \Mathematica, \MATLAB, R, and Python/NumPy, where
all operations are dynamically dispatched (at least semantically, if not
in actual implementation). Such flexibility,
however, has traditionally come at the price of slower execution.
To improve performance, dynamic languages typically resort to static analysis
at some level.
One strategy is to implement arrays in an external library
written in a static language.
The Python NumPy package is a prominent example, implementing array operations
as well as its own type system and internal abstraction mechanisms within
a large C code base \cite{Walt:2011np}. As a result, NumPy \code{ndarray}s are
superficially Python objects, but implementation-wise are disjoint from the
rest of the Python object system, since little of Python's native object
semantics is used to define their behavior.

Another approach is to implement static analyses
\textit{de novo} for dynamic languages. However, the flexibility
of these languages' programs limits the extent of analysis in practice. For example, \MATLAB's array
semantics allow an array to be enlarged automatically whenever a write occurs
to an out-of-bounds index, and also for certain operations to
automatically promote the element type of an array from real to complex
numbers. This poses implementation challenges for static \MATLAB{} compilers
like \code{FALCON}, which have to implement a complete type system with
multiple compiler passes and interprocedural flow analyses to check
for such drastic changes to arrays \cite{Rose:1999tt,
Li:2013mf}. In fact, \MATLAB's (and APL's) semantics are so flexible that shape
inference on arrays is impossible to compute using
ordinary dataflow analysis on bounded lattices \cite{Joisha:2006aa}.
Additionally, type checking is essential to disambiguate
\MATLAB{} expressions like \code{A*B}, which, depending on the dimensions of
\code{A} and \code{B}, could represent a scaling, inner product, outer
product, matrix-matrix multiplication, or matrix-vector multiplication
\cite{Rose:1999tt}. Similar work has been done for other dynamic languages,
as in Hack, a PHP implementation with a full static type system
\cite{Verlaguet:2014hn}.

%TODO other languages: APL, ZPL, Lisp?

The conflicting requirements of performance and flexibility pose a dilemma for
language designers and implementers. Most current languages choose either to
support only programs that are amenable to static analysis for the sake of
performance, like C++ and Haskell, or choose to support more general
classes of programs, like \MATLAB, \Mathematica, and Python/NumPy. While
dynamic languages nominally give up static analysis in the process, many
implementations of these languages still resort to static analysis in
practice, either by hard-coding array semantics \textit{post hoc} in a compiler, or by
implementing arrays in an external library written in a static language.

%A key problem with dynamic languages is how to provide more declarative
%information to a compiler. Users expect to be able to program any
%run-time behavior and have it just work. In theory, a compiler could
%partially evaluate programs based on whatever static information it can
%find (e.g. integer constants), but it is unlikely that this would reliably
%provide performance where it is needed.

%The user should not have to learn two different sets of semantic behavior. In Julia, that's the compiler's job. It does mean that the it's not as fast as a static language.

\begin{figure}
  \centering
  \includegraphics[width=\columnwidth]{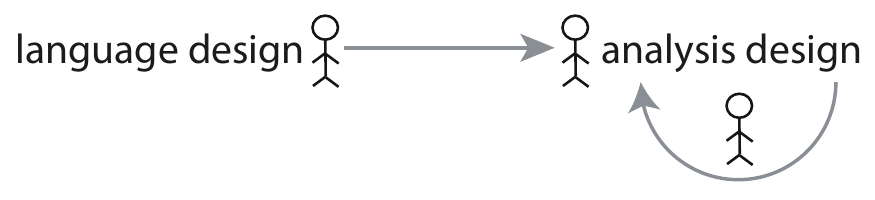}
  \includegraphics[width=\columnwidth]{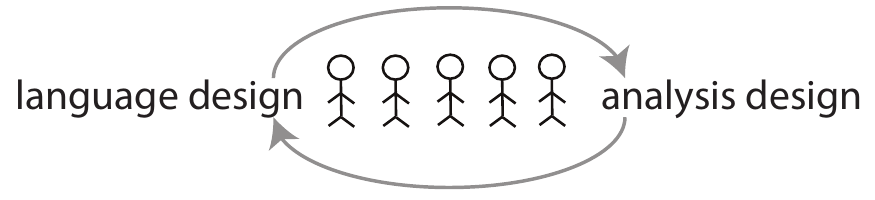}
  \caption{\label{fig:langdesign}
Above: most dynamic languages are designed without consideration for program
analysis, leaving it to future compiler writers if the language becomes
sufficiently popular.
Below: Julia is designed with analysis in mind, with a single community responsible
for both language design and performant implementation. This approach is
natural for statically typed languages, but dynamic languages need static analysis too.}
\end{figure}

\section{Julia arrays}

Julia\cite{Bezanson:2012jf} is dynamically typed and is based on dynamic
multiple dispatch. However, the language and its standard library have been
designed to take advantage of the possibility of static analysis
(Figure~\ref{fig:langdesign}), especially dataflow type inference \cite{Cousot:1977, kaplanullman}. Such type
inference, when combined with multiple dispatch, allows users and library
writers to produce a rich array of specialized methods to handle different
cases performantly. In this section we describe how this language feature
is used to implement indexing for Julia's \code{Array} data type.
The \code{Array} type is parameterized by an element type and a rank (an
integer). For purposes of this paper, its representation can be considered a
tuple of a contiguous memory region and a shape (a tuple of integers giving
the size of the array in each dimension). This simple representation is
already enough to require nontrivial design decisions.

%relevant? Furthermore, the same mechanisms to generate performant code for arrays for native Julia types also extend to user-defined types, since they share the same semantics.

%In the cases where static analysis can happen, Julia can perform as fast as
%languages where array semantics is predicated on static analysis, but multiple
%dispatch allows Julia to be more flexible than just this. Julia also allows
%for generic fall-back methods that are slower because of a lack of static
%analysis, but nonetheless can be dispatched upon. The flexibility that is
%allowed by this approach is a nonobvious application of multiple
%dispatch.

\subsection{Array indexing rules}

Rules must be defined for how various operators act
on array dimensions. Here we will focus on indexing, since selecting parts of
arrays has particularly rich behavior with respect to dimensionality. For
example, if a single row or column of a matrix is selected, does the result
have one or two dimensions? Array implementations prefer to invoke general
rules to answer such questions. Such a rule might say ``dimensions indexed
with scalars are dropped'', or ``trailing dimensions of size one are
dropped'', or ``the rank of the result is the sum of the ranks of the
indexes'' (as in APL \cite{APL}).

The recursive, one-dimension-at-a-time approach favored in static languages limits
which indexing behaviors can be chosen. For example, an indexing expression
of a 3-array in C++ might be written as \code{A[i][j][k]}. Here there are 
three applications of \code{operator[]}, each of which will decide whether to
drop a dimension based on the type of a single index. The second rule
described above, and others like it, cannot be implemented in such a
scheme.

Julia's dispatch mechanism permits a novel approach that encompasses more
rules, and does not require array rank to be known statically, yet
benefits when it is.
This solution is still a compromise among the factors outlined in
the introduction, but it is a new compromise that we find compelling.

%Here we consider the indexing rules. How to compute shapes of subarrays. How
%to deal with singleton dimensions is but a special case even though
%superficially the rules mention them explicitly.

%In Julia, such indexing rules are defined in exactly one place and can be
%changed later if so desired.

\subsection{The need for flexibility}

Our goal here is a bit unusual: we are not concerned with which rules might
work best, but merely with how they can be specified, so that domain experts
can experiment.

In fact, different applications may desire different indexing behaviors. For
example, applications employing arrays with units or other semantic meaning
associated with each dimension may not want to have the dimensions dropped or
rearranged. For example, tomographic imaging applications may want arrays
representing stacks of images as the imaging plane moves through a three-
dimensional object. The resulting array would have associated space/time
dimensions on top of the dimensions indexing into color planes. In
other applications, the dimensions are not semantically distinguishable and it
may be desirable to drop singleton dimensions. For example, a statistical
computation may find it convenient to represent an $n$-point correlation
function in an $n$-array, and integrate over $k$ points to generate the lower
order $(n\!-\!k)$-correlation functions; the indistinguishability of the points
means that the result is most conveniently expressed with rank $(n\!-\!k)$
rather than an $n$-array with $k$ singleton dimensions.

%In linear algebra on the other hand, one might want a row or column of a
% matrix to become a vector by dropping a dimension.

In practice we may have to reach a consensus on what rules to use, but this
should not be forced by technical limitations.

\subsection{Multiple dispatch in Julia}

\begin{figure}
  \centering
  \includegraphics[width=\columnwidth]{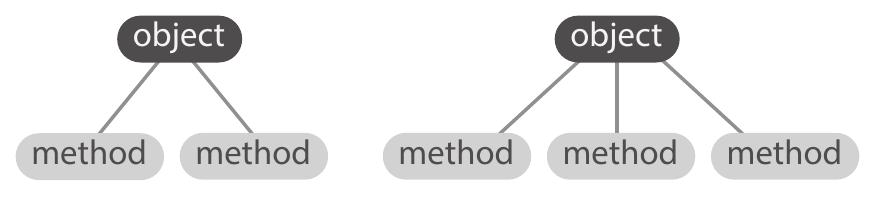}
  \includegraphics[width=\columnwidth]{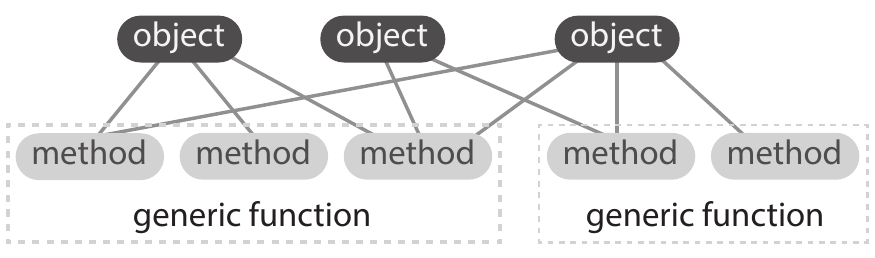}
  \caption{\label{fig:dispatch}Class-based method dispatch (above) vs. multiple dispatch (below).}
\end{figure}

Multiple dispatch (also known as generic functions, or multi-methods) is an
object-oriented paradigm where methods are defined on combinations of data
types (classes), instead of encapsulating methods inside classes
(Figure~\ref{fig:dispatch}). Methods are
grouped into generic functions. A generic function can be applied to several
arguments, and the method with the most specific signature matching the
arguments is invoked.

One can invent examples where multiple dispatch is useful in classic
object-oriented domains such as GUI programming.
For example, a method for drawing a label onto a button might
look like this in Julia syntax:

\begin{minipage}{\linewidth}
\begin{verbatim}
function draw(target::Button, obj::Label)
    ...
end

\end{verbatim}
\end{minipage}

In numerical computing, binary operators are ubiquitous and we can easily imagine
defining special behavior for some combination of two arguments:

\begin{verbatim}
+(x::Real, z::Complex) = complex(x+real(z), imag(z))
\end{verbatim}

But how much more is there? Would we ever need to define a method on
\emph{three} different types at once? Indeed, most language designers and
programmers seem to have concluded that multiple dispatch might be nice, but is
not essential, and the feature is not often used \cite{Muschevici:2008}.
Perhaps the few cases that seem to need it can be handled using tricks like
Python's \code{\_\_add\_\_} and \code{\_\_radd\_\_} methods.

However, in technical computing the need for polymorphic, multi-argument
operators goes further. In fact we have found a need for additional
dispatch features that are not always found in multi-method implementations.
For array semantics, support for variadic methods is perhaps the most
important such feature. Combining multiple dispatch and variadic methods
seems straightforward, yet permits surprisingly powerful definitions, and
entails a surprising amount of complexity. For example,
consider a variadic \code{sum} function that adds up its arguments. We could
write the following two methods for it (note that in Julia, \code{Real} is
the abstract supertype of all real number types, and \code{Integer} is the
abstract supertype of all integer types):

\begin{verbatim}
sum(xs::Integer...)
sum(xs::Real...)
\end{verbatim}

The syntax \code{...} allows an argument slot to match any number of trailing
arguments (currently, Julia only allows this at the end of a method signature).
In the first case, all arguments are integers and so we can use a naive
summation algorithm. In the second case, we know that at least one argument
is not an integer (otherwise the first method would be used), so we might want to use some form of compensated
summation instead. Notice that these modest method signatures
capture a subtle property (at least one argument is non-integral)
\emph{declaratively}, without needing to explicitly loop over the arguments
to examine their types. The signatures also provide useful type information:
at the very least, a compiler could know that all argument values inside
the first method are of type \code{Integer}. Yet the type annotations
are not redundant, but are necessary to specify the desired behavior. There
is also no loss of flexibility, since \code{sum} can be called with any combination
of number types, as users of dynamic technical computing languages would expect.

While the author of these definitions does not write a loop to examine
argument types, such a loop of course still must take place somewhere inside
the dispatch system. Such a complex dispatch system is naturally at risk of
performing badly. However, Julia pervasively applies dataflow type
inference, so that argument types are often known in advance, in turn
allowing method lookup to be done at compile-time. Technically this is
just an optimization, but in practice it has a profound impact on how code
is written.

\subsection{Argument tuple transformations for indexing}

Multiple dispatch appears at first to be about operator overloading:
defining the behavior of functions on new, user-defined types.
But the fact that the compiler ``knows'' the types of function arguments leads
to a surprising, different application: performing elaborate, declarative
transformations of argument tuples.

Determining the result shape of an indexing operation is just such a
transformation. In Julia's standard library, we have a function
\code{index\_shape} that accepts index arguments (which, for present
purposes, may be scalars or arrays of any rank), and returns the
shape (a tuple of integers) of the result. The length of the shape
determines the rank of the result array. Many different behaviors
are possible, but currently we use the rule that trailing dimensions
indexed with scalars are dropped.\footnote{This rule has been the subject of
some debate in the Julia community \cite{issue5949}. Fortunately it is easy to change,
as we will see.}
For example:

\begin{verbatim}
A[1:m, 1:n, 2]     # rank 2
A[1:m, 2, 1:n]     # rank 3
A[1:m, 2, 1:n, 1]  # rank 3
\end{verbatim}

The following two method definitions express this behavior:

{\small
\begin{verbatim}
# drop trailing dimensions indexed with scalars
index_shape(i::Real...) = ()
index_shape(i, I...)    = tuple(length(i),
                                index_shape(I...)...)
\end{verbatim}
}
(The \code{...} ellipsis syntax within an expression, on the right-hand side of
a definition, performs ``argument splicing'': the elements of a container
are spread into multiple arguments to the called function. Formal
arguments that lack a \code{::} type specializer match any value.)
The first definition traps and collapses runs of \code{Real} arguments of
any length. The second definition ensures that the first definition only
applies to the tail of an argument tuple, by keeping indexes as long as
some non-scalar arguments remain.

Since all indexing functions call this function, changing these two lines is
sufficient to change how indexing works. For example, another rule one might
want is to drop \emph{all} dimensions indexed with scalars:

{\small
\begin{verbatim}
# drop dimensions indexed with scalars
index_shape()              = ()
index_shape(i::Real, I...) = index_shape(I...)
index_shape(i, I...)       = tuple(length(i),
                                   index_shape(I...)...)
\end{verbatim}
}

Or we could imitate APL's behavior, where the rank of the result is the sum
of the ranks of the indexes, as follows:

{\small
\begin{verbatim}
# rank summing (APL)
index_shape()        = ()
index_shape(i, I...) = tuple(size(i)...,
                             index_shape(I...)...)
\end{verbatim}
}

Here \code{size} (as opposed to \code{length}) gives the shape tuple of an array,
so we are just concatenating shapes.

\subsection{Exploiting dataflow type inference}

Julia's multi-methods were designed with the idea that dataflow type inference
would be applied to almost all concrete instances of methods, based on
run-time argument types or compile-time estimated argument types.
Our definitions exploit the dataflow operation of matching inferred argument types
against method signatures, thereby destructuring and recurring through argument
tuples at compile-time. As a result, the compiler is able to infer sharp result
types for many variadic calls, and optimize away argument splicing that would otherwise
be prohibitively expensive. More sophisticated method signatures lead to more
sophisticated type deductions.

Tuple types (or product types) are crucial to this analysis. Since the type
of each element of a tuple is tracked, it is possible to deduce that
the type of \code{f(x...)}, where \code{x} has tuple type \code{(A,B)}, is
equal to the type of \code{f} applied to arguments of types \code{A} and
\code{B}. Variadic methods introduce unbounded tuple types, written as
\code{(T...)}. Unbounded tuple types form a lattice of infinite height,
since new subtypes can always be constructed in the sequence
\code{(T...)}, \code{(T,T...)}, \code{(T,T,T...)}, etc. This adds
significant complexity to our lattice operators.

\subsection{Similarities to symbolic pattern matching}

Julia's multi-methods resemble symbolic pattern matching, such as those in
computer algebra systems. Pattern matching systems effectively
allow dispatch on the full structure of values, and so are in some sense
even more powerful than our generic functions. However, they lack a clear
separation between the type and value domains, leading to performance
opacity: it is not clear what the system will be able to optimize
effectively and what it won't.
Such a separation could be addressed by
designating some class of patterns as the ``types'' that the compiler
will analyze. However, more traditional type systems could be seen as
doing this already, while also gaining data abstraction in the bargain.

\subsection{Implications for Julia programmers}

In many array languages, a function like
\code{index\_shape} would be implemented inside the run-time system
(possibly scattered among many functions), and separately embodied in
a hand-written transfer function inside the compiler. Our design shows
that such arrangements can be replaced by a combination of high-level code and
a generic analysis. Similar conclusions on the value of incorporating analyzed library code into a
compiler were drawn by the Telescoping Languages project \cite{telescoping}.
Yet other languages like Single Assignment C allow great flexibility in
user-defined functions but require the built-in shape functions to be
implemented with special purpose type functions
\cite{Scholz:2003sa,Grelck:2006sa}.

From the programmer's perspective, Julia's multi-methods are convenient
because they provide run-time and compile-time abstraction in a single
mechanism. Julia's ``object'' system is also 
its ``template'' system, without different syntax or reasoning about
binding time. Semantically, methods always dispatch on run-time
types, so the same definitions are applicable whether types are known
statically or not. This makes it possible to use popular
dynamic constructs such as \code{A[I...]} where \code{I} is a heterogeneous
array of indexes. In such a case the compiler will need to generate a
dynamic dispatch, but only the performance of the call site is affected.
%Therefore users are free to reason about
%programs operationally, while gaining richer compile-time information
%as their method definitions grow more sophisticated.

One price of this flexibility is that not all such definitions are well-founded:
it is possible to write methods that yield tuples of indeterminate length.
The compiler must recognize this condition and apply widening operators
\cite{Cousot:1977, widening}. In these cases, the deduced types are
still correct but imprecise, and in a way that depends on somewhat arbitrary
choices of widening operators (for example, such a type might look
like \code{(Int...)} or \code{(Int,Int...)}). Nevertheless, we believe that the
flexibility of Julia's multi-methods is of net benefit to programmers.

%This approach does not depend on any heuristics. Each call to
%\texttt{index\_shape} simply requires one recursive invocation of type
%inference. This process reaches the base case \texttt{()} for these
%definitions, since each recursive call handles a shorter argument list (for
%less-well-behaved definitions, we might end up invoking a widening operator
%instead).

%\begin{verbatim}
%diverge() = randbool() ? () : tuple(1, diverge()...)
%\end{verbatim}

%This is an example of indexing behavior that is not amenable to useful static
%analysis, since each branch of \code{diverge()} has different types.

\section{Discussion}

\begin{table}
\label{dispatchratios}
\begin{center}
\begin{tabular}{|l|r|r|r|}\hline
Language & DR & CR & DoS \\
\hline \hline
Gwydion    & 1.74 & 18.27 & 2.14 \\
\hline
OpenDylan  & 2.51 & 43.84 & 1.23 \\
\hline
CMUCL      & 2.03 &  6.34 & 1.17 \\
\hline
SBCL       & 2.37 & 26.57 & 1.11 \\
\hline
McCLIM     & 2.32 & 15.43 & 1.17 \\
\hline
Vortex     & 2.33 & 63.30 & 1.06 \\
\hline
Whirlwind  & 2.07 & 31.65 & 0.71 \\
\hline
NiceC      & 1.36 &  3.46 & 0.33 \\
\hline
LocStack   & 1.50 &  8.92 & 1.02 \\
\hline
Julia      & 5.86 & 51.44 & 1.54 \\
\hline
Julia operators & 28.13 & 78.06 & 2.01 \\
\hline
\end{tabular}
\end{center}
\caption{
Comparison of Julia (1208 functions exported from the \code{Base} library)
to other languages with multiple dispatch.
The ``Julia operators'' row describes 47 functions with special syntax
(binary operators, indexing, and concatenation).
Data for other systems are from Ref.~\cite{Muschevici:2008}.
}
\end{table}

Multiple dispatch is used heavily throughout the Julia ecosystem. To quantify
this statement, we use the following metrics for evaluating the extent of 
multiple dispatch \cite{Muschevici:2008}:

\begin{enumerate}
\item Dispatch ratio (DR): The average number of methods in a generic function.
\item Choice ratio (CR): For each method, the total number of methods over all
generic functions it belongs to, averaged over all methods. This is essentially
the sum of the squares of the number of methods in each generic function, divided
by the total number of methods. The intent of this statistic is to give more weight
to functions with a large number of methods.
\item Degree of specialization (DoS): The average number of type-specialized
arguments per method.
\end{enumerate}

Table~\ref{dispatchratios} shows the mean of each metric over the entire Julia
\code{Base} library, showing a high degree of multiple dispatch compared with
corpora in other languages \cite{Muschevici:2008}.
Compared to most multiple dispatch systems, Julia functions tend to have a large
number of definitions. To see why this might be, it helps to compare results
from a biased sample of only operators. These functions are the most obvious
candidates for multiple dispatch, and as a result their statistics climb
dramatically. Julia is focused on technical computing, and so is likely to
have a large proportion of functions with this character.

\subsection{Other applications}

To give a better sense of how multi-methods
are or could be used in our domain, we will briefly describe a few examples.

\subsubsection{Array views}

In certain instances of array indexing, it is
possible to keep the data in place and return just a view (pointer) to the
data instead of copying it. This functionality is implemented in a Julia
package called \code{ArrayViews.jl}\cite{Lin:2014av}. A crucial property
of an array view is its \emph{contiguous rank}: the number of leading
dimensions for which the strides equal the cumulative product of the shape.
When a view is constructed of another view, the type of view constructed
depends on the indexes used and the contiguous rank of the argument. In
favorable cases, a more efficient \code{ContiguousView} is returned.

This determination is made by a definition similar to the following:

\begin{minipage}{\linewidth}
{\small
\begin{verbatim}
function view(a::DenseArray, I::Subs...)
    shp = vshape(a, I...)
    make_view(a, restrict_crank(contrank(a, I...), shp),
              shp, I...)
end

\end{verbatim}
}
\end{minipage}

\code{contrank} essentially counts the number of leading ``colons'' in the
indexing expression. \code{make\_view} selects (via dispatch) what kind of
view to return based on the result type of \code{restrict\_crank}, which is
set up to return the smaller of its two argument shapes.
This is an excellent example of a library that needs to define behaviors
actually exceeding the complexity of what is provided in the standard library.

\subsubsection{Distributed arrays}

Other classes of array indexing rules are needed in distributed array
implementations. The Star-P system \cite{parry, Choy05parallelmatlab}
let users ``tag'' dimensions as potentially distributed using the notation
\code{*p}, which constructed a special type of object tracked by the system.
Indexing leads to questions of whether to take the first instance of \code{*p}
as the distributed dimension, the last instance, or perhaps just the last dimension.

Such distribution rules could be implemented and experimented with readily
using an approach similar to that used for \code{index\_shape}.

%Though not yet tested, an intriguing thought is that distributed array distribution
%rules can make use of the same approach as general array semantics.
%An interesting example is the  *p syntax used in Star-P. \cite{}
%In this syntax, array dimensions were listed as parallel with the *p notation.
%Julia multiple dispatch syntax provides the flexibility to explore the myriad of choices readily and flexibly.

\subsubsection{Unit quantities}

A package providing unitful computations (\code{SIUnits.jl}\cite{Fischer:2014si})
makes use of the same kinds of
tradeoffs as array semantics. Unitful computations are another case
where the relevent metadata (exponents on units) can be
known at compile-time in many cases, but not always. The
\code{SIUnits} library is free to express the general case, and have the
overhead of tagging and dispatching removed where possible.

The method signatures of operators on unit quantities ensure that they only
apply to arguments with the same units. Because of this design, if an
operator is applied to two arguments where the units are only statically
known for one, the compiler can infer that the other must have the same units
for the operation to succeed. Another implication is that mismatched units can
be handled with a 1-line fallback definition that simply raises an
informative error.

\subsection{Performance}

In this work, we focus on the performance that arises from eliminating
abstraction overhead. Our goal is to convert general definitions (e.g. an
indexing function that can handle many kinds of arguments) into the rough
equivalent of handwritten C code for any particular case. This is a useful
performance target, since programmers often resort to rewriting slow high-
level code in C. Further speedups are possible, but entail techniques beyond
the current scope. Additionally, we reuse the optimization passes provided by
LLVM\cite{LLVM}, allowing us to ignore many lower-level performance issues.

Experience so far suggests that we are close to meeting this
performance goal \cite{Bezanson:2012jf}.

\section{Conclusion}

Programming languages must compromise between the ability to perform
static analyses and allowing maximal flexbility in user programs.
Performance-critical language features like arrays benefit greatly from
static analyses, and so even dynamic languages that initially lack static
analyses eventually want them one way or another.

We speculate that, historically, computer scientists developing multiple 
dispatch were not thinking about technical computing, and those who cared
about technical computing were not interested in the obscurer corners of
object-oriented programming. However, we believe that the combination of
dataflow type inference, sophisticated method signatures, and the need for
high-productivity technical environments is explosive. In this context,
multi-methods, while still recognizable as such, can do work that departs
significantly from familiar uses of operator overloading. By itself, this
mechanism does not address many of the concerns of array programming, such
as memory access order and parallelism. However, we feel it provides a
useful increment of power to dynamic language users who would like to begin
to tackle these and other related problems.

%% multiple
%% dispatch is not merely an optimization hack, but instead can be used to design
%% a programming language where they become core semantic features. The
%% implementation of these features in Julia is a remarkably efficient way to
%% allow code to be specialized to leverage static analyses for performance, yet
%% at the same time allows for other code written for maximal flexibility even in
%% the absence of detailed type information. Static analyzability becomes no
%% longer a property of an entire language, but instead an optional property of
%% specific programs in a particular language. Julia is therefore sufficiently
%% expressive to write complex, generic, dynamic multidimensional behaviors, and
%% more generally allows us to address the performance---flexibility compromise
%% in Julia with greater Pareto optimality than other existing solutions.

%\appendix
%\section{Appendix Title}
%
%This is the text of the appendix, if you need one.

\acks

The authors gratefully acknowledge the enthusiastic participation of the Julia
developer community in many stimulating discussions, in particular Dahua Lin and
Keno Fischer for the \code{ArrayViews.jl}\cite{Lin:2014av} and
\code{SIUnits.jl}\cite{Fischer:2014si} packages, respectively. This
work was supported by the MIT Deshpande Center, an Intel Science and
Technology award, grants from VMWare and Citibank, a Horizontal Software
Fellowship in Compuational Engineering, and NSF DMS-1035400.

% We recommend abbrvnat bibliography style.

\bibliography{refs}{}

\begin{thebibliography}{31}
\providecommand{\natexlab}[1]{#1}
\providecommand{\url}[1]{\texttt{#1}}
\expandafter\ifx\csname urlstyle\endcsname\relax
  \providecommand{\doi}[1]{doi: #1}\else
  \providecommand{\doi}{doi: \begingroup \urlstyle{rm}\Url}\fi

\bibitem[Bavestrelli(2000)]{Bavestrelli:2000ct}
G.~Bavestrelli.
\newblock A class template for {N}-dimensional generic resizable arrays.
\newblock \emph{C/C++ Users Journal}, 18\penalty0 (12):\penalty0 32--43,
  December 2000.
\newblock URL
  \url{http://www.drdobbs.com/a-class-template-for-n-dimensional-gener/184401319}.

\bibitem[Bezanson et~al.(2012)Bezanson, Karpinski, Shah, and
  Edelman]{Bezanson:2012jf}
J.~Bezanson, S.~Karpinski, V.~B. Shah, and A.~Edelman.
\newblock Julia: A fast dynamic language for technical computing.
\newblock arXiv:1209.5145v1, 2012.

\bibitem[Busam and Englund(1969)]{Busam:1969oe}
V.~A. Busam and D.~E. Englund.
\newblock Optimization of expressions in {Fortran}.
\newblock \emph{Communication of the ACM}, 12\penalty0 (12):\penalty0 666--674,
  1969.
\newblock \doi{10.1145/363626.363635}.

\bibitem[Choy and Edelman(2005)]{Choy05parallelmatlab}
R.~Choy and A.~Edelman.
\newblock Parallel {MATLAB}: Doing it right.
\newblock In \emph{Proceedings of the IEEE}, volume~93, pages 331--341, 2005.
\newblock \doi{10.1109/JPROC.2004.840490}.

\bibitem[Cousot and Cousot(1977)]{Cousot:1977}
P.~Cousot and R.~Cousot.
\newblock Abstract interpretation: A unified lattice model for static analysis
  of programs by construction or approximation of fixpoints.
\newblock In \emph{Proceedings of the 4th ACM SIGACT-SIGPLAN Symposium on
  Principles of Programming Languages}, POPL '77, pages 238--252, New York, NY,
  USA, 1977. ACM.
\newblock \doi{10.1145/512950.512973}.

\bibitem[Cousot and Cousot(1992)]{widening}
P.~Cousot and R.~Cousot.
\newblock Comparing the {Galois} connection and widening/narrowing approaches
  to abstract interpretation.
\newblock In M.~Bruynooghe and M.~Wirsing, editors, \emph{Programming Language
  Implementation and Logic Programming}, volume 631 of \emph{Lecture Notes in
  Computer Science}, pages 269--295. Springer Berlin / Heidelberg, 1992.

\bibitem[Falkoff and Iverson(1975)]{APL}
A.~D. Falkoff and K.~E. Iverson.
\newblock The design of {APL}.
\newblock \emph{SIGAPL APL Quote Quad}, 6:\penalty0 5--14, April 1975.
\newblock \doi{10.1145/585923.585925}.

\bibitem[Fischer()]{Fischer:2014si}
K.~Fischer.
\newblock URL \url{https://github.com/loladiro/SIUnits.jl}.

\bibitem[Garcia and Lumsdaine(2005)]{Garcia:2005ma}
R.~Garcia and A.~Lumsdaine.
\newblock {MultiArray}: a {C++} library for generic programming with arrays.
\newblock \emph{Software: Practice and Experience}, 35\penalty0 (2):\penalty0
  159--188, 2005.
\newblock \doi{10.1002/spe.630}.

\bibitem[Grelck and Scholz(2006)]{Grelck:2006sa}
C.~Grelck and S.-B. Scholz.
\newblock {SAC}: A functional array language for efficient multi-threaded
  execution.
\newblock \emph{International Journal of Parallel Programming}, 34\penalty0
  (4):\penalty0 383--427, 2006.
\newblock ISSN 0885-7458.
\newblock \doi{10.1007/s10766-006-0018-x}.

\bibitem[Holy()]{issue5949}
T.~E. Holy.
\newblock Drop dimensions indexed with a scalar?
\newblock URL \url{https://github.com/JuliaLang/julia/issues/5949}.

\bibitem[Husbands(1999)]{parry}
P.~Husbands.
\newblock \emph{Interactive Supercomputing}.
\newblock PhD thesis, Department of Electrical Engineering and Computer
  Science, Massachusetts Institute of Technology, 1999.

\bibitem[Joisha and Banerjee(2006)]{Joisha:2006aa}
P.~G. Joisha and P.~Banerjee.
\newblock An algebraic array shape inference system for {MATLAB}.
\newblock \emph{ACM Transactions on Programming Languages and Systems},
  28\penalty0 (5):\penalty0 848--907, Sept. 2006.
\newblock \doi{10.1145/1152649.1152651}.

\bibitem[Kaplan and Ullman(1980)]{kaplanullman}
M.~A. Kaplan and J.~D. Ullman.
\newblock A scheme for the automatic inference of variable types.
\newblock \emph{Journal of the ACM}, 27\penalty0 (1):\penalty0 128--145,
  January 1980.
\newblock \doi{10.1145/322169.322181}.

\bibitem[Keller et~al.(2010)Keller, Chakravarty, Leshchinskiy, Peyton~Jones,
  and Lippmeier]{Keller:2010rs}
G.~Keller, M.~M. Chakravarty, R.~Leshchinskiy, S.~Peyton~Jones, and
  B.~Lippmeier.
\newblock Regular, shape-polymorphic, parallel arrays in {Haskell}.
\newblock In \emph{Proceedings of the 15th ACM SIGPLAN International Conference
  on Functional Programming}, ICFP '10, pages 261--272, New York, NY, USA,
  2010. ACM.
\newblock \doi{10.1145/1863543.1863582}.

\bibitem[Kennedy et~al.(2001)Kennedy, Broom, Cooper, Dongarra, Fowler, Gannon,
  Johnsson, Mellor-Crummey, and Torczon]{telescoping}
K.~Kennedy, B.~Broom, K.~Cooper, J.~Dongarra, R.~Fowler, D.~Gannon,
  L.~Johnsson, J.~Mellor-Crummey, and L.~Torczon.
\newblock Telescoping languages: A strategy for automatic generation of
  scientific problem-solving systems from annotated libraries.
\newblock \emph{Journal of Parallel and Distributed Computing}, 61\penalty0
  (12):\penalty0 1803 -- 1826, 2001.
\newblock \doi{10.1006/jpdc.2001.1724}.

\bibitem[Lattner and Adve(2004)]{LLVM}
C.~Lattner and V.~Adve.
\newblock {LLVM}: A compilation framework for lifelong program analysis \&
  transformation.
\newblock In \emph{Proceedings of the 2004 International Symposium on Code
  Generation and Optimization (CGO'04)}, pages 75--86, Palo Alto, California,
  Mar 2004.

\bibitem[Li and Hendren(2013)]{Li:2013mf}
X.~Li and L.~Hendren.
\newblock {Mc2For}: a tool for automatically transforming {MATLAB} to {Fortran
  95}.
\newblock Technical Report SABLE-TR-2013-4, Sable Research Group, School of
  Computer Science, McGill University, Montr\'eal, Qu\'ebec, Canada, 2013.
\newblock URL
  \url{http://www.sable.mcgill.ca/publications/techreports/2013-4/techrep.pdf}.

\bibitem[Lin()]{Lin:2014av}
D.~Lin.
\newblock URL \url{https://github.com/lindahua/ArrayViews.jl}.

\bibitem[Lippmeier and Keller(2011)]{Lippmeier:2011ep}
B.~Lippmeier and G.~Keller.
\newblock Efficient parallel stencil convolution in {Haskell}.
\newblock In \emph{Proceedings of the 4th ACM Symposium on Haskell}, Haskell
  '11, pages 59--70, New York, NY, USA, 2011. ACM.
\newblock \doi{10.1145/2034675.2034684}.

\bibitem[Lippmeier et~al.(2012)Lippmeier, Chakravarty, Keller, and
  Peyton~Jones]{Lippmeier:2012gp}
B.~Lippmeier, M.~M.~T. Chakravarty, G.~Keller, and S.~Peyton~Jones.
\newblock Guiding parallel array fusion with indexed types.
\newblock In \emph{Haskell '12 Proceedings of the 2012 Haskell Symposium},
  pages 25--36, New York, 2012. ACM.
\newblock \doi{10.1145/2364506.2364511}.

\bibitem[Liskov and Zilles(1974)]{Liskov:1974pb}
B.~H. Liskov and S.~N. Zilles.
\newblock Programming with abstract data types.
\newblock In \emph{Proceedings of the ACM SIGPLAN symposium on Very high level
  languages}, volume~9, pages 50--59, New York, 1974. ACM.

\bibitem[Muschevici et~al.(2008)Muschevici, Potanin, Tempero, and
  Noble]{Muschevici:2008}
R.~Muschevici, A.~Potanin, E.~Tempero, and J.~Noble.
\newblock Multiple dispatch in practice.
\newblock In \emph{Proceedings of the 23rd ACM SIGPLAN Conference on
  Object-oriented Programming Systems Languages and Applications}, OOPSLA '08,
  pages 563--582, New York, NY, USA, 2008. ACM.
\newblock \doi{10.1145/1449764.1449808}.

\bibitem[Randell and Russell(1964)]{Randell:1964a6}
B.~Randell and L.~J. Russell.
\newblock \emph{{ALGOL 60} Implementation}, volume~5 of \emph{The Automatic
  Programming Information Centre Studies in Data Processing}.
\newblock Academic Press, London, 1964.
\newblock URL
  \url{http://www.softwarepreservation.org/projects/ALGOL/book/Randell_ALGOL_60_Implementation_1964.pdf}.

\bibitem[Rose and Padua(1999)]{Rose:1999tt}
L.~D. Rose and D.~Padua.
\newblock Techniques for the translation of {MATLAB} programs into {Fortran
  90}.
\newblock \emph{ACM Transactions on Programming Languages and Systems},
  21:\penalty0 286--323, 1999.

\bibitem[Sattley(1961)]{Sattley:1961as}
K.~Sattley.
\newblock Allocation of storage for arrays in {ALGOL} 60.
\newblock \emph{Communication of the ACM}, 4\penalty0 (1):\penalty0 60--65,
  1961.
\newblock \doi{10.1145/366062.366088}.

\bibitem[Sattley and Ingerman(1960)]{Sattley:1960as}
K.~Sattley and P.~Z. Ingerman.
\newblock The allocation of storage for arrays in {ALGOL} 60.
\newblock \emph{ALGOL Bulletin}, Sup 13.1:\penalty0 1--15, 1960.

\bibitem[Scholz(2003)]{Scholz:2003sa}
S.-B. Scholz.
\newblock {S}ingle {A}ssignment {C}: Efficient support for high-level array
  operations in a functional setting.
\newblock \emph{Journal of Functional Programming}, 13\penalty0 (6):\penalty0
  1005--1059, 2003.
\newblock \doi{10.1017/S0956796802004458}.

\bibitem[van~der Walt et~al.(2011)van~der Walt, Colbert, and
  Varoquaux]{Walt:2011np}
S.~van~der Walt, S.~C. Colbert, and G.~Varoquaux.
\newblock The {NumPy} array: A structure for efficient numerical computation.
\newblock \emph{Computing in Science \& Engineering}, 13\penalty0 (2):\penalty0
  22--30, 2011.
\newblock \doi{10.1109/MCSE.2011.37}.

\bibitem[Veldhuizen(1998)]{Veldhuizen:1998ab}
T.~L. Veldhuizen.
\newblock Arrays in {Blitz++}.
\newblock In D.~Caromel, R.~R. Oldehoeft, and M.~Tholburn, editors,
  \emph{Computing in Object-Oriented Parallel Environments}, volume 1505 of
  \emph{Lecture Notes in Computer Science}, pages 223--230. Springer Berlin
  Heidelberg, 1998.
\newblock \doi{10.1007/3-540-49372-7_24}.

\bibitem[Verlaguet and Menghrajani(2014)]{Verlaguet:2014hn}
J.~Verlaguet and A.~Menghrajani.
\newblock Hack: a new programming language for {HHVM}.
\newblock March 2014.
\newblock URL \url{http://code.facebook.com/posts/264544830379293}.

\end{thebibliography}
\bibliographystyle{abbrvnat}

% The bibliography should be embedded for final submission.

%\begin{thebibliography}{}
%\softraggedright

%\bibitem[Smith et~al.(2009)Smith, Jones]{smith02}
%P. Q. Smith, and X. Y. Jones. ...reference text...

%\end{thebibliography}

\end{document}